\documentclass[preprint, 
amsmath, amssymb, superscriptaddress, 
nofootinbib, 
preprintnumbers, keywords]{revtex4}

\usepackage{bm}
\usepackage{booktabs}
\usepackage{mathrsfs}

\usepackage[dvipdfmx]{graphicx}
\usepackage{color}
\usepackage{comment}

\definecolor{mygray}{gray}{0.5}
\usepackage{fancybox} 
\usepackage{ulem}
\usepackage{multirow}
\usepackage{here}

\begin{document}

\title{Probing Scalar-Mediator Quark Couplings via CLFV Lepton-Nucleon Scattering}

\author{Yuichiro Kiyo}
\email[E-mail: ]{ykiyo@juntendo.ac.jp}
\affiliation{Department of Physics, Juntendo University,
Inzai, 270-1695, Japan}

\author{Michihisa Takeuchi}
\email[E-mail: ]{takeuchi@mail.sysu.edu.cn}
\affiliation{School of Physics and Astronomy, Sun Yat-sen University, 519082 Zhuhai, China}

\author{Yuichi Uesaka}
\email[E-mail: ]{y-uesaka388@dokkyomed.ac.jp}
\affiliation{Department of Fundamental Education, 
Dokkyo Medical University, 880 Kitakobayashi, Mibu, Shimotsuga, 
Tochigi 321-0293, Japan}

\author{Masato Yamanaka}
\email[E-mail: ]{m.yamanaka.km@cc.it-hiroshima.ac.jp}
\affiliation{Department of Advanced Sciences, 
Faculty of Science and Engineering, 
Hosei University, Tokyo 184-8584, Japan}
\affiliation{Department of Literature, Faculty of Literature, 
Shikoku Gakuin University, Kagawa 765-8505, Japan}
\affiliation{Department of Global Environment Studies, 
Hiroshima Institute of Technology, Hiroshima, 731-5193, Japan}


\date{\today}

\begin{abstract}
{We investigate charged lepton flavor violating (CLFV) deep-inelastic scattering, 
focusing on the gluon-initiated subprocess $\ell_i g \to \ell_j g$ via the gluon 
effective operator $\phi\, G_{\mu \nu}^a G_a^{\mu \nu}$, and demonstrate how to probe the nature of the CLFV 
mediator $\phi$, specifically its mass and interaction with quarks. 
We consider two benchmark scenarios for the mediator–quark coupling: 
(i) $h$-like scenario, in which the mediator couples to heavy quarks in proportion 
to their masses, and (i\hspace{-1pt}i) $b$-only scenario, where the coupling is restricted to bottom quark only. We demonstrate that these scenarios can be discriminated by examining the dependence of the differential cross 
section on the momentum transfer. Furthermore, we show that the peak position of the differential cross section exhibits a pronounced sensitivity to both the mass of the mediator and the coupling strengths
with quarks. 
}
\end{abstract}

\maketitle

\section{Introduction}
\label{sec:Intro}

Charged lepton flavor violating (CLFV) processes provide important tests of
physics beyond the Standard Model (SM). 
Various theoretical models predict interesting 
outcomes, with the rates of relevant CLFV processes 
lying just below current experimental 
bounds~\cite{Calibbi:2017uvl, Bernstein:2013hba, Kuno:1999jp, Raidal:2008jk}. 
On the other hand, sensitivity improvements of a few orders of 
magnitude are expected at upcoming CLFV experiments ~\cite{Adamov:2018vin, Bartoszek:2014mya, Belle-II:2024sce, 
Mu3e:2020gyw}.  Interplay among the different CLFV processes and 
complementary observables can shed light on the flavor structure of the underlying particle physics models.

An interesting and important class of studies involve
CLFV searches in deep-inelastic scattering (DIS) processes, 
$\ell_i N \to \ell_j X$~\cite{Gninenko:2001id, Sher:2003vi, Kanemura:2004jt, 
Gonderinger:2010yn, Bolanos:2012zd, Liao:2015hqu, Abada:2016vzu, Gninenko:2018num, 
Antusch:2020fyz, Husek:2020fru}. Here $\ell_{i(j)}$ denotes  a charged lepton of flavor 
$i (j)$, and $N$ and $X$ denote a nucleon and hadronic final states, respectively. 
The high intensity achieved in lepton-nucleon scattering experiments offers an opportunity to test the particle models beyond the SM.
The expected event rate would exceed $\mathcal{O}(10)$/year for the maximally allowed CLFV couplings at next generation fixed target experiments, where the Avogadro's number enhances the luminosity. The event rate can be enhanced by the incident beam energy. 
So far, the HERA experiment has searched for CLFV DIS events and has set constraints on the relevant 
CLFV parameters~\cite{Aktas:2007ji,Aaron:2011zz}. The sensitivity is expected to improve in upcoming experiments 
such as the EIC~\cite{Accardi:2012qut, Willeke:2021ymc}, LHeC~\cite{LHeC:2020van}, ILC~\cite{ILC:2013jhg, ILCInternationalDevelopmentTeam:2022izu}. 

Among the various models, we focus on (pseudo)scalar CLFV mediators that interact primarily with heavy quarks, as 
they are predicted by several well-motivated particle physics models and are of great theoretical and experimental interest 
(e.g.,~\cite{Agashe:2005hk, Crivellin:2020mjs, Tsumura:2009yf, 
Botella:2015hoa, Crivellin:2013wna, Davidson:2010xv, Kanemura:2005hr, Davidson:2007si, 
Moreau:2006np, Agashe:2006iy,
Huber:2003tu, Cirigliano:2021img, De:2024foq}). 
Measuring the properties of CLFV mediators is challenging at low-energy flavor experiments. Experimental searches for the CLFV decays of leptons provide information only on the effective local interactions involving several matter fields at a space-time point, such as the four-Fermi interaction, and do not allow for the separate determination of the CLFV couplings and mediator mass. 
This limitation can be solved by considering the scattering experiments at high-energy colliders, 
where scattering energy can be varied to disentangle the effects from the CLFV couplings and the mediator mass. 
However, this appoach remains highly challenging at the LHC, where CLFV signals are completely overwhelmed by large QCD backgrounds.
On the other hand, lepton–nucleon scattering experiments would provide a cleaner and more favorable environment for probing CLFV interactions.

This paper discusses how the DIS processes can probe the CLFV interactions mediated by a certain scalar particle. Specifically, we focus on the effective coupling between the mediator and gluons, 
generated through quark loop diagrams (triangle diagrams) and represented by the operator $g_{\phi gg} \phi\, G_{\mu\nu}^a G^{\mu \nu}_a$ (see Fig. 1)~\cite{Takeuchi:2017btl}, where $\phi$ is 
the mediator and $G^{\mu \nu}_a$ is the gluon field strength. The consequent sub-process $\ell_i g \to \ell_j g$ arising from this operator  would be a leading contribution to the CLFV 
DIS~\cite{Takeuchi:2017btl, Husek:2020fru}. 
It should be noted that the effective coupling $g_{\phi gg}$ depends 
on the momentum transfer,  and this dependence is highly sensitive 
to the quark flavors that couple to the mediator. Even under identical 
experimental conditions (e.g., beam energy, beam polarization), the 
differential cross section exhibits different behavior as a function of 
the mediator mass and the flavors of the interacting quarks. As
demonstrated in the subsequent sections, analyzing momentum
distributions across varying beam energies also provides insight into the coupling 
strength between the mediator and quarks. 
The peak position of the differential cross section 
is particularly important, as it can reveal the nature of the CLFV
mediator that remains hidden in low-energy flavor experiments and at the LHC.

This paper is organized as follows: 
In the next section, we introduce the model Lagrangian that describes 
CLFV DIS mediated by a (pseudo-)scalar field, and show the induced
effective operator between the CLFV mediator and gluons.
Then, in Sec.~\ref{Sec:CrossSection}
we calculate the parton level cross section of 
the CLFV subprocess $\ell_i g \to \ell_j g$. 
In Sec.\,\ref{Sec:CouplingDep}, we present a numerical analysis 
illustrating how different coupling scenarios can be 
distinguished through the behavior of the differential cross section.
Sec.~\ref{Sec:Summary} is devoted to the summary and outlook
for the CLFV searches in lepton-nucleon scattering. 

\begin{figure}[H]
\begin{center}
\includegraphics[width=0.4\hsize]{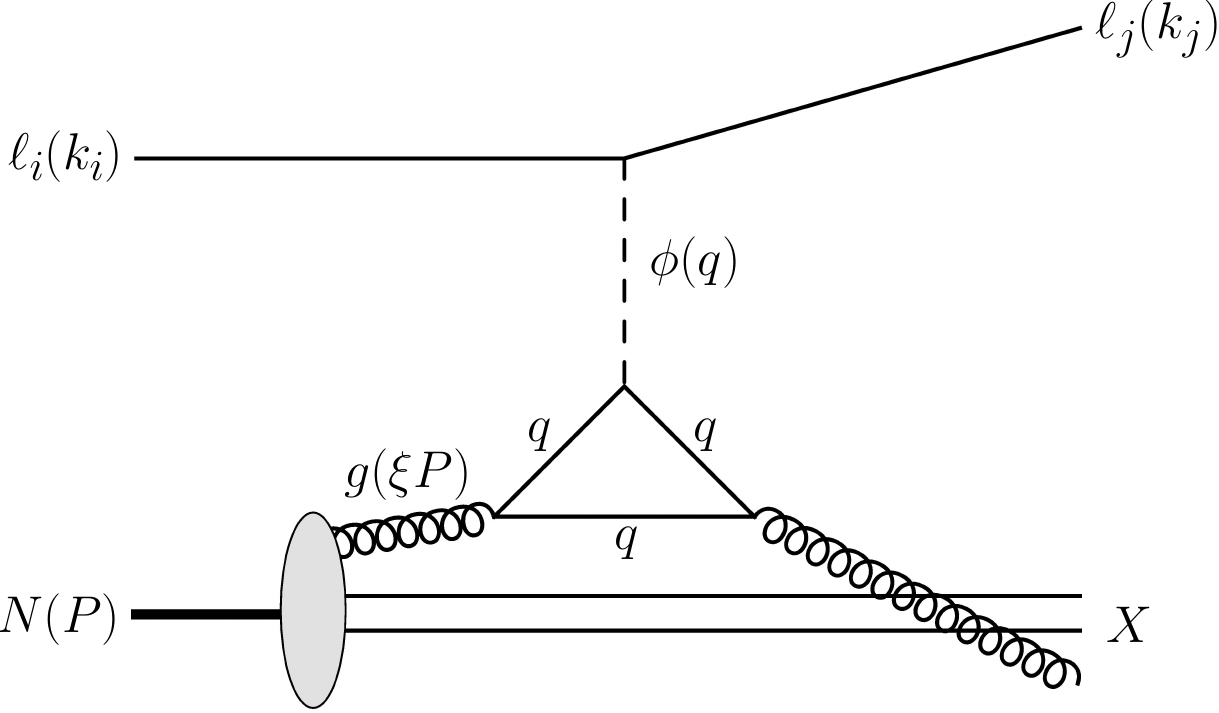}
\end{center}
\vspace{-0.3cm}
\caption{CLFV DIS which corresponds to a subprocess $\ell_i g \to \ell_j g$.
The nucleon momentum is $P$, and a gluon inside nucleon with momentum 
$\xi P$ is scattered by a CLFV mediator $\phi$ which carries a momentum $q=k_i-k_j$. The effective interaction is generated through a triangle loop 
involving heavy quarks $q=c, b, t$.}
\label{fig:LFVDIS}
\end{figure}

\section{CLFV interactions}
\label{Sec:Int}

We introduce a neutral mediator field, denoted as $\phi$, which can be either a scalar ($\phi_S$) or a pseudoscalar ($\phi_P$), and which 
couples to charged leptons $\ell_i$ and heavy-flavor quarks, specifically
 the charm ($c$), bottom  ($b$), and top  ($t$) quarks.
The interaction Lagrangian is given by
\begin{equation}
\begin{split}
	\mathcal{L}_\phi
	&= 
	- \sum_{i, j=e, \mu, \tau} \left( 
	\rho_{ij}^\phi \bar{\ell}_j P_L \ell_i \phi
	+ {\rm h.c.} 
	\right)
	- \sum_{q=c,b,t}\, \rho_{qq}^\phi \bar{q} \Gamma q \phi,
\label{Eq:intL}
\end{split}
\end{equation}
where $i$ and $j$ denote lepton flavor indices, and $q$ represents the heavy quarks.
The projection operator is denoted as $P_L=(1-\gamma_5)/2$, and the Dirac structure is given 
by $\Gamma=1$ for the scalar case and  $\Gamma=i\gamma_5$ for 
the pseudoscalar case. We assume that $\phi$ interacts with the heavy
 quarks through flavor-diagonal couplings $\rho_{qq}^{\phi}$. 
 All fields in this Lagrangian are expressed in the mass eigenstate basis.
For notational simplicity, we denote both the scalar and pseudoscalar couplings
to leptons simply as $\rho^{\phi}_{ij}$, without distinguishing between the two.
When necessary, we explicitly use $\rho_{ij}^S$ and $\rho_{ij}^P$ to refer to the 
scalar and pseudoscalar cases, respectively. 
The same convention is applied to quark couplings.

In the following, we consider two patterns of the mediator-quark couplings.
The first case is the $h$-like scenario, where the coupling strength $\rho_{qq}^\phi$ 
is proportional to the corresponding quark mass:
\begin{eqnarray}
\rho_{qq}^\phi &=&  \kappa^\phi  \frac{m_q}{v} 
~~~~ \mbox{[$h$-like scenario]},
\label{Eq:h-like_scenario}
\end{eqnarray}
where $m_q$ is the mass of the heavy quark $q$, and $v=(\sqrt{2}G_F)^{-1/2} = 246\, {\rm GeV}$ 
is the vacuum expectation value of the SM Higgs field.  
With this normalization, $\kappa_q^\phi=1$ corresponds to the case where $\rho_{qq}^\phi$ is equal to the SM Higgs coupling to heavy quarks.

The second is a scenario in which the mediator couples to only one type of heavy quark, but not to the others.
We refer to this as the $q$-only scenario.
For instance, the $b$-only scenario is defined as
\begin{eqnarray}
\rho_{bb}^\phi=\rho_{(b)}^\phi ,~~
\rho_{cc}^\phi= \rho_{tt}^\phi=0~~\mbox{[$b$-only scenario]},
\label{Eq:q-only_scenario}
\end{eqnarray}
where $\rho_{(b)}^\phi$ is non-zero, and the mediator couples
 selectively to the bottom quark but not to charm or top quarks. Similarly, $c$-only  and $t$-only scenarios can be defined with $\rho_{(c)}^\phi$ and $\rho_{(t)}^\phi$,
respectively, but we do not discuss them further in 
this paper.

Although the mediator does not couple directly to gluons, 
an interaction with gluons is induced at the one-loop level via a triangle diagram
with heavy quarks running in the loop, as shown in Fig.~\ref{fig:LFVDIS}.
The corresponding effective interaction Lagrangian for the scalar mediator $\phi_S$ 
is given by 
\begin{eqnarray}
\mathcal{L}_{\phi gg}
	&=&
	g_{Sgg}\,  \phi_{S}\,  G_{\mu \nu}^{a} G^{\mu \nu}_a,
	\label{Eq:SGG}
\end{eqnarray}
where $G_{\mu\nu}$ denotes the gluon field strength tensor. 
For the pseudoscalar mediator $\phi_P$,
the effective interaction is given by
\begin{eqnarray}
\mathcal{L}_{\phi gg}
	&= & 
	g_{Pgg}\, \phi_{P}\,  G_{\mu \nu}^{a} \widetilde{G}^{\mu \nu}_a, 
	\label{Eq:PGG}
\end{eqnarray}
where $\widetilde{G}_{\mu \nu}^{a}= \frac{1}{2}\epsilon_{\mu\nu \rho \sigma} \, 
 G^{a\rho \sigma}$ is the dual field strength tensor of the gluon. 
The coupling constants $g_{Sgg}$ and $g_{Pgg}$ are obtained by evaluating the one-loop triangle diagrams. 
Their explicit forms are well known in the literature ~\cite{Georgi:1977gs, Spira:1995rr}, and are given by 
\begin{eqnarray}
g_{Sgg} 
	&=&
	\sum_{q=c, b, t} \frac{\alpha_{s} \rho_{qq}^{S}}{8\pi m_{q}} \, 
	\tau_{q} \Bigl[ 1+\left( 1-\tau_{q} \right) 
	f\left( \tau_{q} \right) \Bigr],
\label{Eq:g_Sgg}
\\
g_{Pgg} 
	&=& 
	\sum_{q=c,b,t} \frac{\alpha_{s} \rho_{qq}^{P}}{8\pi m_{q}} \, 
	\tau_{q} f\left( \tau_{q} \right), 
\label{Eq:g_Pgg}
\end{eqnarray}
where  $\alpha_s=g_s^2/(4\pi)$
 is the strong coupling constant of QCD, and $\tau_{q}= 4m_{q}^{2}/q^{2}$, with $q^2$ being the squared momentum of the scalar or pseudoscalar field. 
 The one-loop function $f \left( \tau \right)$ is defined as 
\begin{eqnarray}
	f \left( \tau \right) = - \frac{1}{4} 
	\ln^{2} 
	\left[ 
	- \, \frac{1+\sqrt{1-\tau}}{1-\sqrt{1-\tau}}
	\right] 
	~~~~ (\tau<0). 
\label{Eq:ftau}
\end{eqnarray}
In the DIS process shown in Fig.~\ref{fig:LFVDIS}, the CLFV mediator is exchanged in the $t$-channel. Since the momentum transfer is spacelike ($q^2<0$),
 the loop-function $f \left( \tau \right)$ has
  no imaginary part  in this kinematical region.

\begin{figure}[t!]
\centering
\label{a}\includegraphics[width=.49\linewidth]{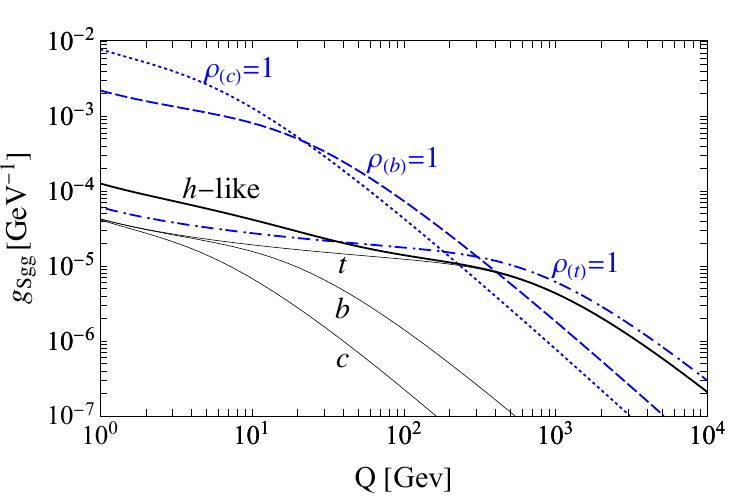}
\label{b}\includegraphics[width=.49\linewidth]{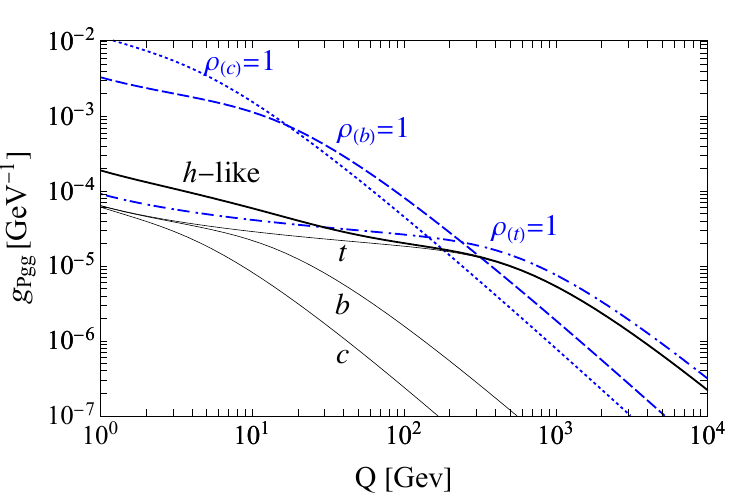}
\caption{$Q$-dependence of the mediator-gluon coupling for scalar (left panel) 
and pseudoscalar (right panel) cases. 
The solid line corresponds to  $h$-like scenario with $\kappa^{\phi}=1$, while
the dotted, dashed and dot-dashed lines represent the $c$-only, $b$-only and $t$-only scenarios with coupling constants $\rho_{(c)}^\phi = 1$, $\rho_{(b)}^\phi = 1$ and $\rho_{(t)}^\phi = 1$, respectively.}
\label{Fig:phigg}
\end{figure}

In Fig.~\ref{Fig:phigg}, we show the $Q$-dependence ($Q=\sqrt{-q^2}$) of the scalar and pseudoscalar couplings, $g_{Sgg}$ and $g_{Pgg}$, 
displayed in the left and right panels, respectively, for the $h$-like and $q$-only scenarios.
The input parameters are taken as $m_c=1.3$\,GeV,  $m_b=4.75$\,GeV, and $m_t=172$\,GeV~\cite{Zyla:2020zbs}. 
The solid line corresponds to the $h$-like scenario with $\kappa^\phi=1$, and 
individual flavor contributions from top, bottom, and charm quarks  (labeled as $t, b, c$) 
to this scenario are also shown.
For $q$-only scenarios, the $c$-only, $b$-only and $t$-only cases with 
$\rho_{(c)}^\phi=1, \rho_{(b)}^\phi=1$ and $\rho_{(t)}^\phi=1$ 
are shown by the dotted, dashed, and dot-dashed lines, respectively. 
The  $t$-only scenario with $\rho_{(t)}^\phi=1$ yields a result close to
 the top-quark contribution  in the $h$-like scenario, 
 since the top-quark Yukawa coupling is of order one ($m_t/v\simeq 0.7$).

The behavior of $g_{\phi gg}$ in the low-$Q$ 
limit (equivalently, large-$\tau_q$) can be understood from the approximation $f(\tau_q)\approx \frac{1}{\tau_q}+\frac{1}{3\tau_q^2}$ 
for large $\tau_q$. This leads to $\tau_q\left[1+\left(1-\tau_q\right)f(\tau_q)\right]\approx 2/3$ 
for the scalar coupling, and $\tau_q f(\tau_q)\approx 1$ for the pseudoscalar coupling as $Q^2\to 0$. 
Therefore, the mediator-gluon coupling at low $Q$ $(Q\ll 2m_c)$ can be approximated as
\begin{eqnarray}
g_{\phi gg}(0) &=& \frac{\alpha_s c_\phi}{8\pi}\,  
\left(\frac{\rho_{cc}^{\phi} }{m_c}+\frac{\rho_{bb}^{\phi} }{m_b}+\frac{\rho_{tt}^{\phi} }{m_t}
\right),
\label{Eq:mediator_gluon_coupling_low-Q}
\end{eqnarray}
with $c_\phi=\frac{2}{3}$ and $1$ for the scalar and pseudoscalar cases, respectively. 
Thus, each quark contribution affects only the overall normalization of $g_{\phi gg}$,
and low-energy measurements, such as $\tau \to e \pi^+ \pi^-$ and 
$\tau \to e \pi^0$, of this coupling cannot resolve the individual 
quark contributions to the mediator-gluon interaction.

When $Q\gg m_q$, the loop function can be approximated as  $f(\tau_q)\approx -\frac{1}{4} \ln^2\frac{Q^2}{m_q^2}$.
The leading high-$Q$ behavior
of the mediator-gluon coupling in the $q$-only scenario is then given by
\begin{eqnarray}
g_{\phi gg}^{q\text{-only}}(Q)
&\approx &
\frac{\alpha_s(Q)}{8\pi}
\left(\frac{\rho_{(q)}^\phi}{m_q}\right)\,
\frac{m_q^2}{Q^2}
\ln^2\left(\frac{Q^2}{m_q^2}\right)
\hspace{1cm} (2m_q\ll Q),
\label{Eq:mediator_gluon_coupling_high-Q}
\end{eqnarray}
where the strong coupling constant $\alpha_s(\mu)$ is evaluated at the scale $\mu=Q$.

Using  Eqs.~ \eqref{Eq:mediator_gluon_coupling_low-Q} and  \eqref{Eq:mediator_gluon_coupling_high-Q}, 
the $Q$-dependence of the mediator-gluon coupling can be understood. 
Assuming a quark mass hierarchy $m_c\ll m_b \ll m_t$ for simplicity, the 
$Q$-dependence of $g_{\phi gg}$ in the $h$-like scenario can be approximated
as
\begin{eqnarray}
g_{\phi gg}^{h\text{-like}}(Q) 
&\approx &
\frac{\alpha_s(Q)\kappa^\phi}{8\pi v}
\left\{
\begin{array}{ll}
2c_\phi +\frac{m_c^2}{Q^2}\, \ln^2\left(\frac{Q^2}{m_c^2}\right) & (2m_c\ll Q\ll 2m_b)
\\
c_\phi + \frac{m_b^2}{Q^2}\, \ln^2\left(\frac{Q^2}{m_b^2}\right) & (2m_b\ll Q\ll 2m_t)
\\
~~~~~
 \frac{m_t^2}{Q^2}\, \ln^2\left(\frac{Q^2}{m_t^2}\right) & (2m_t\ll Q)
\end{array}
\right. .
\label{Eq:mediator_gluon_coupling_high-Q_h}
\end{eqnarray}
This result shows that a measurement of the $Q$-dependence of $g_{\phi gg}$ 
can reveal the pattern of heavy-quark couplings to the mediator as a function of the energy scale $Q$.

\section{Cross section}
\label{Sec:CrossSection} 

As shown in Fig.~\ref{fig:LFVDIS}, we assign momenta $k_i$ and $k_j$
to the initial- and final-state leptons, respectively, and express the gluon momentum inside the nucleon as $\xi P$, where $\xi$ is the momentum
  fraction and $P$ ($P^2\approx 0$) is the nucleon momentum. 
  The hadronic cross section $\sigma$
  for lepton-nucleon scattering 
$\ell_i  N \to \ell_j X$ (where $X$ denotes the hadronic system in the final state) 
is given by  the convolution of the partonic cross section 
$\hat{\sigma}$ with parton distribution function (PDF) as 
\begin{eqnarray}
\bigg[ \frac{d\sigma}{dxdy} \bigg] \left(  \ell_i N \rightarrow \ell_j X\right) 
&=&
\int_0^1 d\xi  
\bigg[ \frac{d\hat{\sigma}}{dxdy}  \bigg] \left(\ell_i g \to \ell_j g \right) f_{g/N}(\xi,\mu_f),
\end{eqnarray}
where $x =Q^2/(2P\cdot q)$ and $y=P\cdot q/P\cdot k_i$, and $q=k_i-k_f$ 
with $Q^2=-q^2$. 
The function $f_{g/N}(\xi, \mu_f)$ is the gluon PDF, which describes the probability 
that a gluon carries a momentum fraction $\xi$ inside the nucleon at the factorization scale $\mu_f$.
 
With the mediator-gluon interactions of Eqs.~\eqref{Eq:SGG} and \eqref{Eq:PGG}, 
one can then calculate the partonic cross section for the process $\ell_i g\to \ell_j g$. 
To analyze $Q$-dependence of the cross section,
it is convenient to consider the differential cross section with respect to $Q$.
By performing a variable transformation from $y$ to $Q$ and integrating over $x$, 
one obtains the differential cross section as 
\begin{eqnarray}
\bigg[\frac{d\sigma}{dQ}\bigg](\ell_i N\to \ell_j X) 
 &=&
 \frac{|g_{\phi gg}(Q)|^2}{4\pi s^2} \frac{Q^5 L_{ij}^\phi}{(Q^2+m_\phi^2)^2} 
 \times \mathcal{M}_g(Q^2),
 \label{eq:xs_dsdQ}
\end{eqnarray}
with $s=(P+k_i)^2$. 
The function $\mathcal{M}_g$ is the  inverse moment of the gluon PDF, defined by
\begin{eqnarray}
\mathcal{M}_g(Q^2) &=&
\int_{x_{\rm min}}^1 x^{-2} f_{g/N}(x, Q)dx,
\label{eq:inverse_moment}
\end{eqnarray}
where $x_\text{min}= Q^2/s$, and 
the factorization scale $\mu_f$ is 
set equal to $Q$. 
The leptonic part $L_{ij}^\phi$ is given by
\begin{eqnarray}
L_{ij}^\phi
&=&
	\left( |\rho_{ij}^\phi|^2 +|\rho_{ji}^\phi|^2\right)
	\left(Q^2+m_i^2+m_j^2\right)
	+4\, {\rm Re} \left( \rho_{ij}^\phi \rho_{ji}^\phi\right) m_i m_j,
\label{Eq:leptonic_part}
\end{eqnarray}
and arises from the CLFV mediator-lepton interaction given in Eq.~\eqref{Eq:intL}.

\section{Sensitivity to quark coupling scenarios} 
\label{Sec:CouplingDep}

We analyze the CLFV cross sections for electron-proton scattering,
$e P\to \tau X$, mediated by the partonic subprocess $eg \to \tau g$. 
The heavy quark masses are  taken to be the same as those in Sec.~\ref{Sec:Int}.
The $\tau$-lepton mass is set to $m_\tau=1.78~{\rm GeV}$, and the electron mass is neglected.
In this section, we focus on the case where the mediator is a scalar ($\phi=\phi_S$), 
while the pseudoscalar case can be treated in a similar manner.
The CLFV coupling is chosen to match the projected experimental sensitivity 
for the branching ratio $\mathrm{Br}(\tau \to e\pi^+\pi^-) = 5 \times 10^{-10}$~\cite{Belle-II:2018jsg}. 
According to the calculation of low-energy CLFV decay rates%
\footnote{
Translating the result of Ref.~\cite{Celis:2014asa} into our notation, 
the branching ratios for the processes arising from mediator-gluon interactions are given by
$\mathrm{Br}(\tau \to e \pi^+ \pi^-) =
2.6 \times 10^5 \times A_{\phi_S} ~[\mathrm{GeV}^6]$ 
and 
$\mathrm{Br}(\tau \to e \eta') = 2.9 \times 10^6~ A_{\phi_P}~[\mathrm{GeV}^6]$
with 
$A_{\phi_X} \equiv \big|\frac{\overline{\rho}_{\tau e}^{\phi_X}}{m_\phi^2}
\sum_{q} \frac{\rho_{qq}^{\phi_X}}{m_q}\big|^2$.
}
based on Ref.~\cite{Celis:2014asa}, the magnitudes of the coupling 
constants corresponding to the expected sensitivity are estimated 
for the $h$-like and $b$-only scenarios as follows:
\begin{eqnarray}
|\kappa^\phi \, \overline{\rho}_{\tau e}^{\phi}|
 &=& 3.6 \times \left(\frac{m_\phi}{1~\mathrm{TeV}}\right)^2
\hspace{1cm}
(\mbox{$h$-like scenario}),
\label{eq:constraint_on_CLFV_coupling_h}
\\
|\rho_{(b)}^\phi \, \overline{\rho}_{\tau e}^{\phi}|
 &=& 0.21 \times \left(\frac{m_\phi}{1~\mathrm{TeV}}\right)^2
\hspace{1cm}
(\mbox{$b$-only scenario}),
\label{eq:constraint_on_CLFV_coupling_b}
\end{eqnarray}
where 
$\overline{\rho}_{\tau e}^\phi = \sqrt{|\rho_{\tau e}^{\phi}|^2 + |\rho_{e \tau}^{\phi}|^2}$.
We note that in each scenario the coupling is taken to be proportional to $m_\phi^2$ in 
order to keep the CLFV decay rates fixed.

\begin{figure}[htbp]
\centering
\label{a}\includegraphics[width=.6\linewidth]{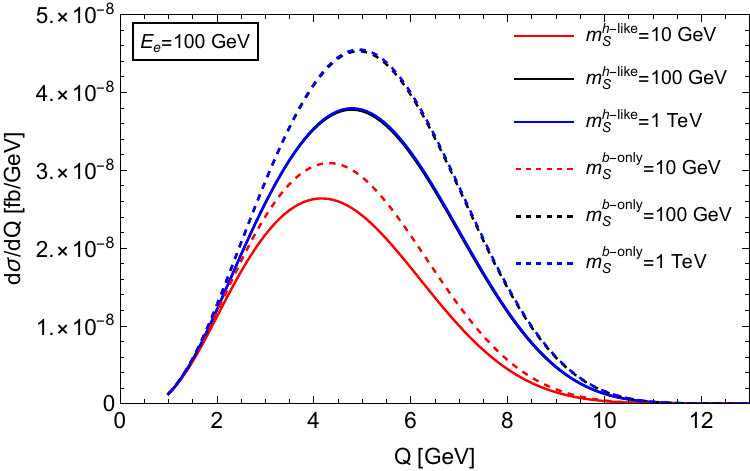}
\\\vspace{3mm}
\label{b}\includegraphics[width=.6\linewidth]{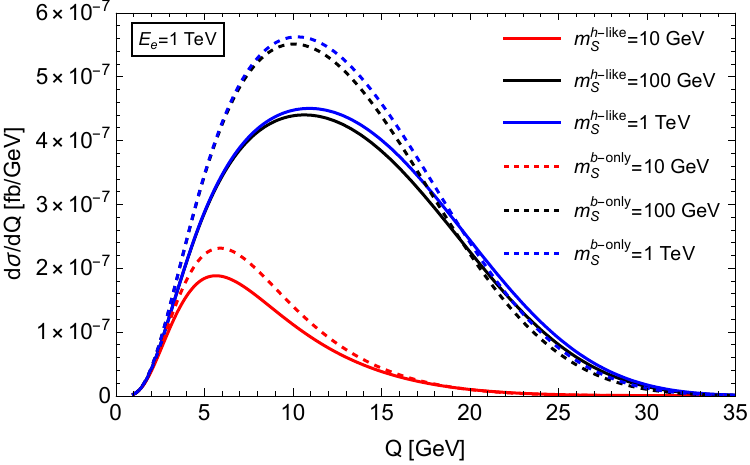}
\\\vspace{3mm}
\label{c}\includegraphics[width=.6\linewidth]{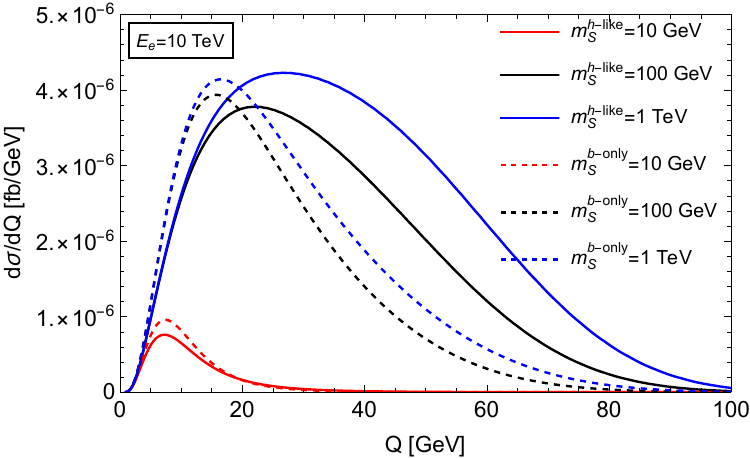}
\caption{Differential cross sections for the process $eg \to \tau g$ as functions of $Q$, 
shown for electron beam energies of $E_e = 100$\,GeV (left panel), 
1\,TeV (middle panel) and 10\,TeV (right panel).  The CLFV coupling $\overline{\rho}_{\tau e}^\phi$
 is set to match the projected sensitivity of $\text{BR}(\tau \to e \pi^+ \pi^-) = 5 \times 10^{-10}$ 
 at a future experiment. Solid (dashed) curves correspond to $h$-like ($b$-only) scenario, 
with mediator masses of $m_S=100$\,GeV (red), 1\,TeV (black), and 10\,TeV (blue). 
}
\label{Fig:DifXS-tau2epipi}
\end{figure}

In Fig.~\ref{Fig:DifXS-tau2epipi}, the differential cross section for electron-proton 
scattering via the partonic process $eg \to \tau g$ is shown. 
We use the \text{CT14 LO} PDF set  ~\cite{Dulat:2015mca} for the proton. 
The electron beam energy is taken to be $E_e = 100~\mathrm{GeV}$ (top panel), 
$1~\mathrm{TeV}$ (middle panel), and $10~\mathrm{TeV}$ (bottom panel).
The solid lines represent the cross sections for the $h$-like scenario, 
while the dashed lines correspond to the $b$-only scenario. 
The scalar mediator mass is varied as $m_\phi = 10~\mathrm{GeV}$ (red), 
$100~\mathrm{GeV}$ (black), and $1~\mathrm{TeV}$ (blue). 
All the lines converge around $Q\sim 1$~GeV because the couplings are chosen according 
to the Eqs.~\eqref{eq:constraint_on_CLFV_coupling_h} and \eqref{eq:constraint_on_CLFV_coupling_b} to reproduce the same branching ratio 
$\mathrm{Br}(\tau \to e\pi^+\pi^-) = 5 \times 10^{-10}$, which is determined using the effective gluon-gluon mediator coupling at $Q \sim 1$~GeV.

In Fig.~\ref{Fig:DifXS-tau2epipi}, the relative size of the differential cross section 
in the $b$-only scenario with respect to that in the $h$-like scenario is greater than 
unity for $Q \lesssim 20$ GeV, but falls below unity for $Q \gtrsim 20$ GeV.
The ratio of the differential 
cross section in the $b$-only scenario to that in the $h$-like scenario becomes 
the ratio of the coupling constants:
$(\overline{\rho}_{\tau e}^{b\text{-only}} g^{b\text{-only}}_{\phi gg}
(Q))^2/(\overline{\rho}_{\tau e}^{h\text{-like}}g^{h\text{-like}}_{\phi gg}(Q))^2$.
This ratio increases and peaks at $Q=8$ GeV, because in the low $Q$ region the effective coupling in the $h$-like scenario drops more rapidly than that in the $b$-only scenario due to the fast decrease of the charm-quark contribution. 
Then it keeps decreasing as a function of $Q$, crossing unity at $Q\simeq 20$ GeV, where the effective coupling in the $h$-like scenario becomes larger than that of $b$-only scenario.

In the top panel ($E_e = 100~\mathrm{GeV}$), 
when the scalar mass is as low as $10~\mathrm{GeV}$, 
the cross section varies with the scalar mass in both scenarios.
In contrast, when the scalar mass is large compared to the typical $Q$ values, 
which are of order $Q \sim \sqrt{2M E_e x} \ \leq 14~\mathrm{GeV}$ for $E_e=100~\mathrm{GeV}$ and the nucleon mass $M\simeq 1\,\text{GeV}$, 
the cross section becomes only weakly dependent on the scalar mass. As a result, the cross sections 
for $m_S = 100~\mathrm{GeV}$ and $m_S = 1~\mathrm{TeV}$ are  degenerate in both scenarios.
This behavior can be understood 
from the relative propagator factor $1/(1 + Q^2/m_S^2)^2$, 
which leads to stronger suppression at high $Q$ for  smaller scalar masses.

 At this beam energy, the peak position of the cross section, denoted as $Q=Q^\ast$, is nearly identical 
between the two scenarios ($Q^\ast = 4 \sim 5~\mathrm{GeV}$), but $h$-like scenario gives a slightly smaller value.
This is because charm-quark contribution in the 
$h$-like scenario decreases rapidly with increasing $Q$, 
shiftig the peak position $Q^\ast$ to a lower value.
In both scenarios, the peak position $Q^\ast$ shifts to smaller values for smaller $m_S$, 
as the propagator suppression becomes more significant at high $Q$ for lighter mediators.

In the middle panel ($E_e = 1~\mathrm{TeV}$), slight differences between 
$m_S=100~\mathrm{GeV}$ and $m_S=1~\mathrm{TeV}$ begin to appear
in the $h$-like and the $b$-only scenarios. At this beam energy, the peak position 
of the cross section for the light scalar mass $m_S = 10~\mathrm{GeV}$
 is around $Q^\ast \sim 6~\mathrm{GeV}$ for both scenarios,  while for intermediate
 and heavy  scalar masses ($m_S = 100~\mathrm{GeV}$ to $1~\mathrm{TeV}$)
the peak shifts to higher values of $Q^\ast \sim 10~\mathrm{GeV}$. 
As is shown in Fig.~\ref{Fig:phigg},
for the same scalar mass $m_S = 10~\mathrm{GeV}$ the peak position $Q^*$ 
for the $h$-like scenario is smaller compared with that for the $b$-only scenario due to the rapid drop of the charm-quark contribution. 
On the other hand, for $m_S=100$~GeV and 1~TeV, the peak position $Q^*$ 
for the $h$-like scenario gives a larger value compared with the $b$-only scenario, 
since the peak position is already around $Q^* \sim 10$~GeV and the effect of the lasting top-quark contribution is more important.

In the bottom panel ($E_e = 10~\mathrm{TeV}$) the 
magnitude of the  
cross section strongly depends on 
the mediator mass $m_S$. In addition,
 for heavy mediator masses, the peak position of the cross section 
 becomes sensitive to the difference between the $h$-like and $b$-only 
 scenarios through the difference of the top-quark contribution, which is a notable feature at this beam energy. 
 For instance, the peak position for $m_S = 100~\mathrm{GeV}$
    ($1~\mathrm{TeV}$) is $Q^* \sim 22~\mathrm{GeV}$ ($27~\mathrm{GeV}$) 
 in the $h$-like scenario, and $Q^* \sim 15~\mathrm{GeV}$ ($16~\mathrm{GeV}$)
 in the $b$-only scenario.

 \begin{table}[!htbp]
\caption{The differential cross section $\frac{d\sigma}{dQ}$ evaluated at the peak position $Q=Q^\ast$ 
for $h\text{-like}$ and $b$-only scenarios. The model parameters
for mediator-quark coupling and the CLFV interaction are taken as $\kappa^\phi=1$, 
$\rho^\phi_{(b)}=1$ and $\bar{\rho}^\phi_{e \tau} = 1$. 
}
\label{table:DiffXS}
\centering
\begin{tabular}{c|c|cc|cc}
\hline
$E_e$
&~ $m_S$~[GeV] ~
& \multicolumn{2}{c|}{$h$-like} & \multicolumn{2}{c}{$b$-only}  \\
&~ & $Q^\ast$~[GeV] & ~$~\frac{d\sigma^h}{dQ}~{\rm [fb/GeV]}$~ 
&~ $Q^\ast$~[GeV] & ~$~\frac{d\sigma^b}{dQ}~{\rm [fb/GeV]}$~
\\  \hline \hline
             & $10$~  & 4.08 & $2.01 \times 10^{-1}$ & 4.24 & $2.65 \times 10^{-2}$    \\
$100~\mathrm{GeV}$ & $10^2$  &  4.78  &  $2.85 \times 10^{-5}$  & 4.88 & $3.86 \times 10^{-6}$ \\
             & $10^3$  &  4.78  &  $2.87 \times 10^{-9}$ & 4.88 & $3.88 \times 10^{-10}$\\
\hline
              & $10$~  & 5.60 &  1.41\hspace{1.35cm} & 5.83 & $1.95 \times 10^{-1}$ \\
$1~\mathrm{TeV}$  & $10^2$  &  10.6  &  $3.29 \times 10^{-4}$ & 9.95 & $4.61 \times 10^{-5}$\\
              & $10^3$  &  10.8  &  $3.36 \times 10^{-8}$ & 10.1 & $4.71 \times 10^{-9}$\\
\hline
               & $10$~  & 7.24  & 5.71\hspace{1.35cm} & 7.39 & $8.05 \times 10^{-1}$\\
$10~\mathrm{TeV}$   & $10^2$  &  22.0  &  $2.80 \times 10^{-3}$ & 15.4 & $3.27 \times 10^{-4}$\\
               & $10^3$  &  26.8  &  $3.13 \times 10^{-7}$ & 16.3 & $3.44 \times 10^{-8}$\\ 
  \hline
\end{tabular}
\end{table}

\begin{figure}[!htbp]
\centering
\label{a}\includegraphics[width=.6\linewidth]{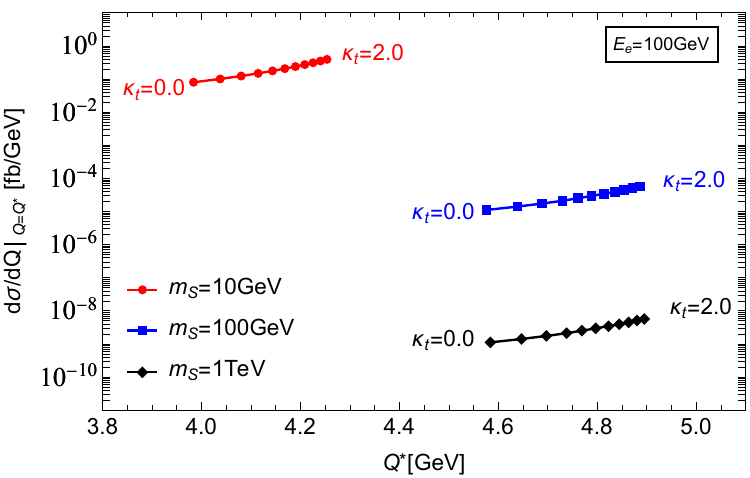}
\\ \vspace{3mm}
\label{b}\includegraphics[width=.6\linewidth]{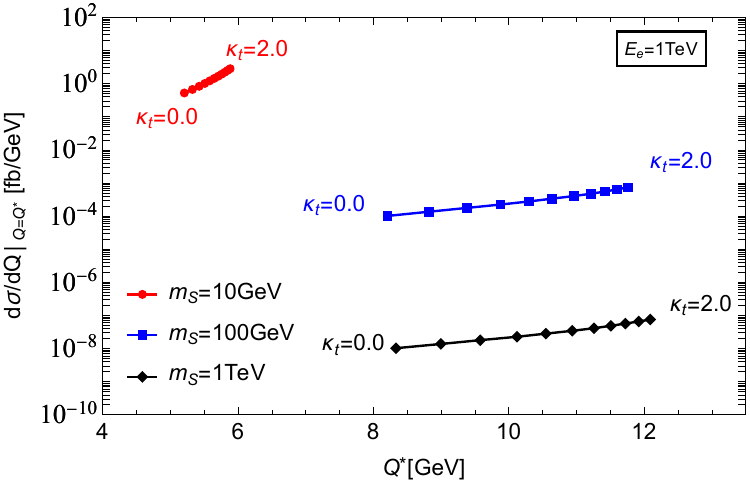}
\\ \vspace{3mm}
\label{c}\includegraphics[width=.6\linewidth]{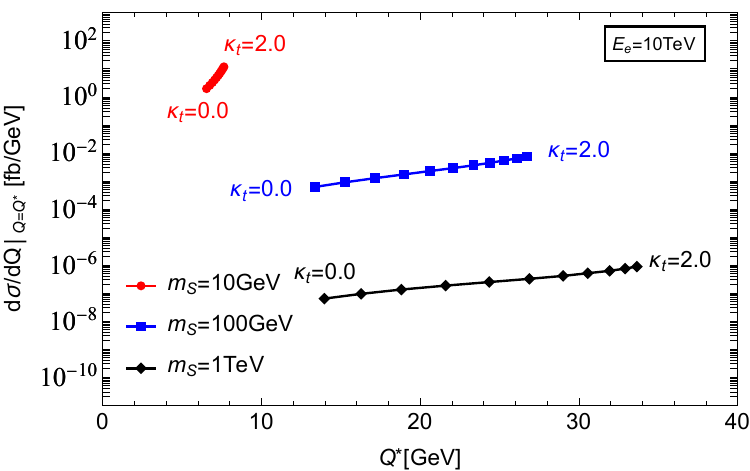}
\caption{
Differential cross section $\frac{d\sigma}{dQ}$ evaluated at the peak position  $Q^\ast$ 
 varying the value of $\kappa_t$ from $0$ to $2$ with an increment of $0.2$.
The electron beam energy is taken to be $E_e=100~{\rm GeV}$(top panel), $1~{\rm TeV}$(middle panel) and $10~{\rm TeV}$(bottom panel). 
The scalar mass is set to $m_S=10~{\rm GeV}$(red), $100~{\rm GeV}$ (blue) and $1~{\rm TeV}$(black).
}
\label{Fig:DifXS-kt}
\end{figure}


Table~\ref{table:DiffXS} summarizes the peak positions $Q^*$ and the corresponding magnitude of the 
differential cross sections at these peak positions for the $h$-like and $b$-only scenarios, with electron beam energies $E_e = 100~\mathrm{GeV}, 1~\mathrm{TeV}$, and $10~\mathrm{TeV}$,  and scalar mediator masses $m_S = 10$, $10^2$, and $10^3~\mathrm{GeV}$.
In the table, the CLFV coupling is fixed to  $\overline{\rho}_{\tau e}^\phi = 1$, 
with $\kappa^\phi = 1$ and $\rho^\phi_{(b)}=1$.\footnote{
To obtain the cross section values shown in Fig.~\ref{Fig:DifXS-tau2epipi}, 
multiply the values in Table~\ref{table:DiffXS} by 
$|\kappa^\phi\overline{\rho}_{\tau e}^\phi|^2$ 
or $|\rho_{(b)}^\phi \overline{\rho}_{\tau e}^\phi|^2$,
 as given respectively in 
Eqs.~\eqref{eq:constraint_on_CLFV_coupling_h} and 
\eqref{eq:constraint_on_CLFV_coupling_b}.}
For $E_e = 100~\mathrm{GeV}$ and $1~\mathrm{TeV}$, the peak positions in the two scenarios are close to each other. However at $E_e = 10~\mathrm{TeV}$, 
the difference becomes noticeable: 
$Q^*|_{\text{$h$-like}} - Q^*|_{\text{$b$-only}} \sim 7~\mathrm{GeV}$ for $m_\phi = 100~\mathrm{GeV}$, and 
$Q^*|_{\text{$h$-like}} - Q^*|_{\text{$b$-only}}\sim 10~\mathrm{GeV}$ for $m_\phi = 1~\mathrm{TeV}$.
This difference originates from the $Q$-dependence of the mediator-gluon coupling 
$g_{\phi gg}(Q)$, as shown in Eqs.~\eqref{Eq:mediator_gluon_coupling_high-Q}
and~\eqref{Eq:mediator_gluon_coupling_high-Q_h}.

Determining the peak position in CLFV DIS experiments may allow one to extract information about the coupling between the mediator and heavy quarks,  
which arises through one-loop triangle diagrams, 
as given in Eqs.~\eqref{Eq:g_Sgg} and~\eqref{Eq:g_Pgg}. 
This approach of probing the $Q$-dependence in high-$Q$ DIS experiments provides a promising means to access model details that cannot be determined from low-energy CLFV searches, such as the $\tau$-decay experiments.

To further demonstrate the sensitivity of the cross section to the mediator-quark coupling, 
we introduce an alternative coupling scenario:  
\begin{eqnarray}
\rho_{cc}^\phi = \frac{m_c}{v}, ~~~
\rho_{bb}^\phi = \frac{m_b}{v},~~~
\rho_{tt}^\phi = \kappa_t^\phi \frac{m_t}{v},
\end{eqnarray}
where $\kappa_t^\phi$ parametrizes the enhancement factor of the mediator's coupling
 to the top quark, and $\kappa_t^\phi = 1$ corresponds to the $h$-like scenario. 
Fig.~\ref{Fig:DifXS-kt} shows how the peak position $Q^*$ and the magnitude of the 
cross section at the peak vary as $\kappa_t^\phi$ is increased from 0 to 2 in increments of 0.2.  
Here, the CLFV coupling constant is fixed to $\overline{\rho}_{\tau e}=1$.
The top, middle, and bottom panels correspond to the electron beam energies of  
$E_e = 100~\mathrm{GeV}, 1~\mathrm{TeV}$, and $10~\mathrm{TeV}$, respectively.  
The red, blue, and black markers indicate the scalar masses of 
$m_S = 10~\mathrm{GeV}, 100~\mathrm{GeV}$, and $1~\mathrm{TeV}$, respectively.  
A clear sensitivity of the peak position $Q^*$ to the mediator-top-quark coupling 
$\kappa_t^\phi$ is observed. 
In the top panel ($E_e=100~{\rm GeV}$),  
peak position $Q^\ast$ changes by only about 0.3~GeV as $\kappa_t^\phi$ varied from 0 to 2 for all $m_S=10, 100$~GeV, and 1~TeV.
In the bottom panel ($E_e=10~{\rm TeV}$), 
the peak position 
$Q^\ast$ varies from $14$ to $34$~GeV for $m_S = 1$~TeV
while the shift of $Q^*$ is smaller for $m_S=10$~GeV.
In general, increasing the beam energy enhances the  sensitivity of the coupling owing to the broader $Q^2$ coverage, although this improvement is less pronounced for lighter mediator masses.

\section{Summary}
\label{Sec:Summary} 

In this paper, 
we have carried out a numerical analysis of CLFV DIS events mediated by the gluon-initiated process through the effective coupling $g_{Sgg}(Q)$.
To demonstrate how different mediator-quark interaction structures can be probed, we have analyzed in detail the two representative coupling scenarios:
(i) the $h$-like scenario, where the mediator couples to heavy quarks in a manner analogous to the SM Higgs, i.e., with strength proportional to their masses, $\rho_{qq} \propto m_q$; and 
(i\hspace{-1pt}i) the $b$-only scenario, 
where the mediator couples only to bottom quarks. 

First, we showed the $Q^2$ dependence of the effective gluon coupling, emphasizing the differences between the two scenarios. The discrepancy in the differential cross sections between the two scenarios originates from the difference in the $Q^2$ dependence of the effective coupling. 
We show that, due to the rapid decrease of the charm-quark contribution and slow decrease of the top-quark contribution, the effective coupling $g_{Sgg}(Q)$ in the $h$-like scenario decreases faster than that for $b$-only scenario at low $Q$, whereas opposite behavior is observed at relatively high $Q$.

Next, we show the differential cross section for several mediator masses for both scenarios. The lighter mediator mass shifts the charm-quark contribution position to lower values in both scenarios due to the propagator suppression. We also discuss how to distinguish the two scenarios using the peak position $Q^\ast$. 
Even for the scenarios sharing the same effective coupling strength at very low $Q$, they can still be discriminated by examining the $Q$-dependence of the differential cross sections.

Finally, we have demonstrated that the 
parameter $\kappa_t$ 
in the modified $h$-like scenario can be determined for various mediator masses by measuring the peak position and 
magnitude of the differential cross section.
The results indicate that increasing the beam energy enhances the sensitivity to $\kappa_t$ values, especially for relatively heavy mediator masses.
For lighter mediators, however, the improvement is limited due to the propagator suppression at high $Q$.

While this paper focuses solely on the differential cross section of the gluon-initiated process $e g\to \tau g$ to examine the effects of the mediator mass and coupling scenarios, additional independent information could be obtained by analyzing differential cross sections of other processes, such as heavy-quark-associated production channels like $e g \to b \bar{b} \tau$. A detailed investigation of such processes is 
beyond the scope of the present work, but will be explored in future studies.

\section*{Acknowledgements}

This work is supported in part by the JSPS Grant-in-Aid for
Scientific Research Numbers 
No.~JP22K03602 (Y.K., Y.U., and M.Y.)
No.~JP25K07325 (Y.K.), 
No.~JP23K13106 (Y.U.), 
No.~20H05852 and No.~22K03638 (M.Y.).
M.T. is supported by the Fundamental Research Funds for the Central Universities, the One Hundred Talent Program of Sun Yat-sen University, China,
and by the JSPS KAKENHI Grant, the Grant-in-Aid for Scientific Research\,C, Grant No.~18K03611.
This work was partly supported by MEXT 
Joint Usage/Research Center on Mathematics and Theoretical 
Physics JPMXP0619217849. 

\bibliography{Ref-eptotauX}

\begin{thebibliography}{45}
\expandafter\ifx\csname natexlab\endcsname\relax\def\natexlab#1{#1}\fi
\expandafter\ifx\csname bibnamefont\endcsname\relax
  \def\bibnamefont#1{#1}\fi
\expandafter\ifx\csname bibfnamefont\endcsname\relax
  \def\bibfnamefont#1{#1}\fi
\expandafter\ifx\csname citenamefont\endcsname\relax
  \def\citenamefont#1{#1}\fi
\expandafter\ifx\csname url\endcsname\relax
  \def\url#1{\texttt{#1}}\fi
\expandafter\ifx\csname urlprefix\endcsname\relax\def\urlprefix{URL }\fi
\providecommand{\bibinfo}[2]{#2}
\providecommand{\eprint}[2][]{\url{#2}}

\bibitem[{\citenamefont{Calibbi and Signorelli}(2018)}]{Calibbi:2017uvl}
\bibinfo{author}{\bibfnamefont{L.}~\bibnamefont{Calibbi}} \bibnamefont{and}
  \bibinfo{author}{\bibfnamefont{G.}~\bibnamefont{Signorelli}},
  \bibinfo{journal}{Riv. Nuovo Cim.} \textbf{\bibinfo{volume}{41}},
  \bibinfo{pages}{71} (\bibinfo{year}{2018}), \eprint{1709.00294}.

\bibitem[{\citenamefont{Bernstein and Cooper}(2013)}]{Bernstein:2013hba}
\bibinfo{author}{\bibfnamefont{R.~H.} \bibnamefont{Bernstein}}
  \bibnamefont{and} \bibinfo{author}{\bibfnamefont{P.~S.}
  \bibnamefont{Cooper}}, \bibinfo{journal}{Phys. Rept.}
  \textbf{\bibinfo{volume}{532}}, \bibinfo{pages}{27} (\bibinfo{year}{2013}),
  \eprint{1307.5787}.

\bibitem[{\citenamefont{Kuno and Okada}(2001)}]{Kuno:1999jp}
\bibinfo{author}{\bibfnamefont{Y.}~\bibnamefont{Kuno}} \bibnamefont{and}
  \bibinfo{author}{\bibfnamefont{Y.}~\bibnamefont{Okada}},
  \bibinfo{journal}{Rev. Mod. Phys.} \textbf{\bibinfo{volume}{73}},
  \bibinfo{pages}{151} (\bibinfo{year}{2001}), \eprint{hep-ph/9909265}.

\bibitem[{\citenamefont{Raidal et~al.}(2008)}]{Raidal:2008jk}
\bibinfo{author}{\bibfnamefont{M.}~\bibnamefont{Raidal}} \bibnamefont{et~al.},
  \bibinfo{journal}{Eur. Phys. J.} \textbf{\bibinfo{volume}{C57}},
  \bibinfo{pages}{13} (\bibinfo{year}{2008}), \eprint{0801.1826}.

\bibitem[{\citenamefont{Abramishvili et~al.}(2018)}]{Adamov:2018vin}
\bibinfo{author}{\bibfnamefont{R.}~\bibnamefont{Abramishvili}}
  \bibnamefont{et~al.} (\bibinfo{collaboration}{COMET}) (\bibinfo{year}{2018}),
  \bibinfo{note}{[PTEP2020,no.3,033C01(2020)]}, \eprint{1812.09018}.

\bibitem[{\citenamefont{Bartoszek et~al.}(2014)}]{Bartoszek:2014mya}
\bibinfo{author}{\bibfnamefont{L.}~\bibnamefont{Bartoszek}}
  \bibnamefont{et~al.} (\bibinfo{collaboration}{Mu2e}),
  \bibinfo{journal}{arXiv}  (\bibinfo{year}{2014}), \eprint{1501.05241}.

\bibitem[{\citenamefont{Adachi et~al.}(2024)}]{Belle-II:2024sce}
\bibinfo{author}{\bibfnamefont{I.}~\bibnamefont{Adachi}} \bibnamefont{et~al.}
  (\bibinfo{collaboration}{Belle-II}) (\bibinfo{year}{2024}),
  \eprint{2405.07386}.

\bibitem[{\citenamefont{Arndt et~al.}(2021)}]{Mu3e:2020gyw}
\bibinfo{author}{\bibfnamefont{K.}~\bibnamefont{Arndt}} \bibnamefont{et~al.}
  (\bibinfo{collaboration}{Mu3e}), \bibinfo{journal}{Nucl. Instrum. Meth. A}
  \textbf{\bibinfo{volume}{1014}}, \bibinfo{pages}{165679}
  (\bibinfo{year}{2021}), \eprint{2009.11690}.

\bibitem[{\citenamefont{Gninenko et~al.}(2002)\citenamefont{Gninenko, Kirsanov,
  Krasnikov, and Matveev}}]{Gninenko:2001id}
\bibinfo{author}{\bibfnamefont{S.~N.} \bibnamefont{Gninenko}},
  \bibinfo{author}{\bibfnamefont{M.~M.} \bibnamefont{Kirsanov}},
  \bibinfo{author}{\bibfnamefont{N.~V.} \bibnamefont{Krasnikov}},
  \bibnamefont{and} \bibinfo{author}{\bibfnamefont{V.~A.}
  \bibnamefont{Matveev}}, \bibinfo{journal}{Mod. Phys. Lett.}
  \textbf{\bibinfo{volume}{A17}}, \bibinfo{pages}{1407} (\bibinfo{year}{2002}),
  \eprint{hep-ph/0106302}.

\bibitem[{\citenamefont{Sher and Turan}(2004)}]{Sher:2003vi}
\bibinfo{author}{\bibfnamefont{M.}~\bibnamefont{Sher}} \bibnamefont{and}
  \bibinfo{author}{\bibfnamefont{I.}~\bibnamefont{Turan}},
  \bibinfo{journal}{Phys. Rev.} \textbf{\bibinfo{volume}{D69}},
  \bibinfo{pages}{017302} (\bibinfo{year}{2004}), \eprint{hep-ph/0309183}.

\bibitem[{\citenamefont{Kanemura et~al.}(2005)\citenamefont{Kanemura, Kuno,
  Kuze, and Ota}}]{Kanemura:2004jt}
\bibinfo{author}{\bibfnamefont{S.}~\bibnamefont{Kanemura}},
  \bibinfo{author}{\bibfnamefont{Y.}~\bibnamefont{Kuno}},
  \bibinfo{author}{\bibfnamefont{M.}~\bibnamefont{Kuze}}, \bibnamefont{and}
  \bibinfo{author}{\bibfnamefont{T.}~\bibnamefont{Ota}},
  \bibinfo{journal}{Phys. Lett.} \textbf{\bibinfo{volume}{B607}},
  \bibinfo{pages}{165} (\bibinfo{year}{2005}), \eprint{hep-ph/0410044}.

\bibitem[{\citenamefont{Gonderinger and
  Ramsey-Musolf}(2010)}]{Gonderinger:2010yn}
\bibinfo{author}{\bibfnamefont{M.}~\bibnamefont{Gonderinger}} \bibnamefont{and}
  \bibinfo{author}{\bibfnamefont{M.~J.} \bibnamefont{Ramsey-Musolf}},
  \bibinfo{journal}{JHEP} \textbf{\bibinfo{volume}{11}}, \bibinfo{pages}{045}
  (\bibinfo{year}{2010}), \bibinfo{note}{[Erratum: JHEP 05, 047 (2012)]},
  \eprint{1006.5063}.

\bibitem[{\citenamefont{Bolanos et~al.}(2013)\citenamefont{Bolanos, Fernandez,
  Moyotl, and Tavares-Velasco}}]{Bolanos:2012zd}
\bibinfo{author}{\bibfnamefont{A.}~\bibnamefont{Bolanos}},
  \bibinfo{author}{\bibfnamefont{A.}~\bibnamefont{Fernandez}},
  \bibinfo{author}{\bibfnamefont{A.}~\bibnamefont{Moyotl}}, \bibnamefont{and}
  \bibinfo{author}{\bibfnamefont{G.}~\bibnamefont{Tavares-Velasco}},
  \bibinfo{journal}{Phys. Rev. D} \textbf{\bibinfo{volume}{87}},
  \bibinfo{pages}{016004} (\bibinfo{year}{2013}), \eprint{1212.0904}.

\bibitem[{\citenamefont{Liao and Wu}(2016)}]{Liao:2015hqu}
\bibinfo{author}{\bibfnamefont{W.}~\bibnamefont{Liao}} \bibnamefont{and}
  \bibinfo{author}{\bibfnamefont{X.-H.} \bibnamefont{Wu}},
  \bibinfo{journal}{Phys. Rev. D} \textbf{\bibinfo{volume}{93}},
  \bibinfo{pages}{016011} (\bibinfo{year}{2016}), \eprint{1512.01951}.

\bibitem[{\citenamefont{Abada et~al.}(2017)\citenamefont{Abada, De~Romeri,
  Orloff, and Teixeira}}]{Abada:2016vzu}
\bibinfo{author}{\bibfnamefont{A.}~\bibnamefont{Abada}},
  \bibinfo{author}{\bibfnamefont{V.}~\bibnamefont{De~Romeri}},
  \bibinfo{author}{\bibfnamefont{J.}~\bibnamefont{Orloff}}, \bibnamefont{and}
  \bibinfo{author}{\bibfnamefont{A.}~\bibnamefont{Teixeira}},
  \bibinfo{journal}{Eur. Phys. J. C} \textbf{\bibinfo{volume}{77}},
  \bibinfo{pages}{304} (\bibinfo{year}{2017}), \eprint{1612.05548}.

\bibitem[{\citenamefont{Gninenko et~al.}(2018)\citenamefont{Gninenko,
  Kovalenko, Kuleshov, Lyubovitskij, and Zhevlakov}}]{Gninenko:2018num}
\bibinfo{author}{\bibfnamefont{S.}~\bibnamefont{Gninenko}},
  \bibinfo{author}{\bibfnamefont{S.}~\bibnamefont{Kovalenko}},
  \bibinfo{author}{\bibfnamefont{S.}~\bibnamefont{Kuleshov}},
  \bibinfo{author}{\bibfnamefont{V.~E.} \bibnamefont{Lyubovitskij}},
  \bibnamefont{and} \bibinfo{author}{\bibfnamefont{A.~S.}
  \bibnamefont{Zhevlakov}}, \bibinfo{journal}{Phys. Rev. D}
  \textbf{\bibinfo{volume}{98}}, \bibinfo{pages}{015007}
  (\bibinfo{year}{2018}), \eprint{1804.05550}.

\bibitem[{\citenamefont{Antusch et~al.}(2020)\citenamefont{Antusch, Hammad, and
  Rashed}}]{Antusch:2020fyz}
\bibinfo{author}{\bibfnamefont{S.}~\bibnamefont{Antusch}},
  \bibinfo{author}{\bibfnamefont{A.}~\bibnamefont{Hammad}}, \bibnamefont{and}
  \bibinfo{author}{\bibfnamefont{A.}~\bibnamefont{Rashed}}
  (\bibinfo{year}{2020}), \eprint{2003.11091}.

\bibitem[{\citenamefont{Husek et~al.}(2021)\citenamefont{Husek, Monsalvez-Pozo,
  and Portoles}}]{Husek:2020fru}
\bibinfo{author}{\bibfnamefont{T.}~\bibnamefont{Husek}},
  \bibinfo{author}{\bibfnamefont{K.}~\bibnamefont{Monsalvez-Pozo}},
  \bibnamefont{and} \bibinfo{author}{\bibfnamefont{J.}~\bibnamefont{Portoles}},
  \bibinfo{journal}{JHEP} \textbf{\bibinfo{volume}{01}}, \bibinfo{pages}{059}
  (\bibinfo{year}{2021}), \eprint{2009.10428}.

\bibitem[{\citenamefont{Aktas et~al.}(2007)}]{Aktas:2007ji}
\bibinfo{author}{\bibfnamefont{A.}~\bibnamefont{Aktas}} \bibnamefont{et~al.}
  (\bibinfo{collaboration}{H1}), \bibinfo{journal}{Eur. Phys. J. C}
  \textbf{\bibinfo{volume}{52}}, \bibinfo{pages}{833} (\bibinfo{year}{2007}),
  \eprint{hep-ex/0703004}.

\bibitem[{\citenamefont{Aaron et~al.}(2011)}]{Aaron:2011zz}
\bibinfo{author}{\bibfnamefont{F.}~\bibnamefont{Aaron}} \bibnamefont{et~al.}
  (\bibinfo{collaboration}{H1}), \bibinfo{journal}{Phys. Lett. B}
  \textbf{\bibinfo{volume}{701}}, \bibinfo{pages}{20} (\bibinfo{year}{2011}),
  \eprint{1103.4938}.

\bibitem[{\citenamefont{Accardi et~al.}(2016)}]{Accardi:2012qut}
\bibinfo{author}{\bibfnamefont{A.}~\bibnamefont{Accardi}} \bibnamefont{et~al.},
  \bibinfo{journal}{Eur. Phys. J. A} \textbf{\bibinfo{volume}{52}},
  \bibinfo{pages}{268} (\bibinfo{year}{2016}), \eprint{1212.1701}.

\bibitem[{\citenamefont{Willeke}(2021)}]{Willeke:2021ymc}
\bibinfo{author}{\bibfnamefont{F.}~\bibnamefont{Willeke}}
  (\bibinfo{year}{2021}).

\bibitem[{\citenamefont{Agostini et~al.}(2021)}]{LHeC:2020van}
\bibinfo{author}{\bibfnamefont{P.}~\bibnamefont{Agostini}} \bibnamefont{et~al.}
  (\bibinfo{collaboration}{LHeC, FCC-he Study Group}), \bibinfo{journal}{J.
  Phys. G} \textbf{\bibinfo{volume}{48}}, \bibinfo{pages}{110501}
  (\bibinfo{year}{2021}), \eprint{2007.14491}.

\bibitem[{ILC(2013)}]{ILC:2013jhg}
 (\bibinfo{year}{2013}), \eprint{1306.6352}.

\bibitem[{\citenamefont{Aryshev
  et~al.}(2022)}]{ILCInternationalDevelopmentTeam:2022izu}
\bibinfo{author}{\bibfnamefont{A.}~\bibnamefont{Aryshev}} \bibnamefont{et~al.}
  (\bibinfo{collaboration}{ILC International Development Team})
  (\bibinfo{year}{2022}), \eprint{2203.07622}.

\bibitem[{\citenamefont{Agashe et~al.}(2005)\citenamefont{Agashe, Papucci,
  Perez, and Pirjol}}]{Agashe:2005hk}
\bibinfo{author}{\bibfnamefont{K.}~\bibnamefont{Agashe}},
  \bibinfo{author}{\bibfnamefont{M.}~\bibnamefont{Papucci}},
  \bibinfo{author}{\bibfnamefont{G.}~\bibnamefont{Perez}}, \bibnamefont{and}
  \bibinfo{author}{\bibfnamefont{D.}~\bibnamefont{Pirjol}}
  (\bibinfo{year}{2005}), \eprint{hep-ph/0509117}.

\bibitem[{\citenamefont{Crivellin et~al.}(2021)\citenamefont{Crivellin, Greub,
  M\"uller, and Saturnino}}]{Crivellin:2020mjs}
\bibinfo{author}{\bibfnamefont{A.}~\bibnamefont{Crivellin}},
  \bibinfo{author}{\bibfnamefont{C.}~\bibnamefont{Greub}},
  \bibinfo{author}{\bibfnamefont{D.}~\bibnamefont{M\"uller}}, \bibnamefont{and}
  \bibinfo{author}{\bibfnamefont{F.}~\bibnamefont{Saturnino}},
  \bibinfo{journal}{JHEP} \textbf{\bibinfo{volume}{02}}, \bibinfo{pages}{182}
  (\bibinfo{year}{2021}), \eprint{2010.06593}.

\bibitem[{\citenamefont{Tsumura and Velasco-Sevilla}(2010)}]{Tsumura:2009yf}
\bibinfo{author}{\bibfnamefont{K.}~\bibnamefont{Tsumura}} \bibnamefont{and}
  \bibinfo{author}{\bibfnamefont{L.}~\bibnamefont{Velasco-Sevilla}},
  \bibinfo{journal}{Phys. Rev. D} \textbf{\bibinfo{volume}{81}},
  \bibinfo{pages}{036012} (\bibinfo{year}{2010}), \eprint{0911.2149}.

\bibitem[{\citenamefont{Botella et~al.}(2016)\citenamefont{Botella, Branco,
  Nebot, and Rebelo}}]{Botella:2015hoa}
\bibinfo{author}{\bibfnamefont{F.}~\bibnamefont{Botella}},
  \bibinfo{author}{\bibfnamefont{G.}~\bibnamefont{Branco}},
  \bibinfo{author}{\bibfnamefont{M.}~\bibnamefont{Nebot}}, \bibnamefont{and}
  \bibinfo{author}{\bibfnamefont{M.}~\bibnamefont{Rebelo}},
  \bibinfo{journal}{Eur. Phys. J. C} \textbf{\bibinfo{volume}{76}},
  \bibinfo{pages}{161} (\bibinfo{year}{2016}), \eprint{1508.05101}.

\bibitem[{\citenamefont{Crivellin et~al.}(2013)\citenamefont{Crivellin, Kokulu,
  and Greub}}]{Crivellin:2013wna}
\bibinfo{author}{\bibfnamefont{A.}~\bibnamefont{Crivellin}},
  \bibinfo{author}{\bibfnamefont{A.}~\bibnamefont{Kokulu}}, \bibnamefont{and}
  \bibinfo{author}{\bibfnamefont{C.}~\bibnamefont{Greub}},
  \bibinfo{journal}{Phys. Rev. D} \textbf{\bibinfo{volume}{87}},
  \bibinfo{pages}{094031} (\bibinfo{year}{2013}), \eprint{1303.5877}.

\bibitem[{\citenamefont{Davidson and Grenier}(2010)}]{Davidson:2010xv}
\bibinfo{author}{\bibfnamefont{S.}~\bibnamefont{Davidson}} \bibnamefont{and}
  \bibinfo{author}{\bibfnamefont{G.~J.} \bibnamefont{Grenier}},
  \bibinfo{journal}{Phys. Rev. D} \textbf{\bibinfo{volume}{81}},
  \bibinfo{pages}{095016} (\bibinfo{year}{2010}), \eprint{1001.0434}.

\bibitem[{\citenamefont{Kanemura et~al.}(2006)\citenamefont{Kanemura, Ota, and
  Tsumura}}]{Kanemura:2005hr}
\bibinfo{author}{\bibfnamefont{S.}~\bibnamefont{Kanemura}},
  \bibinfo{author}{\bibfnamefont{T.}~\bibnamefont{Ota}}, \bibnamefont{and}
  \bibinfo{author}{\bibfnamefont{K.}~\bibnamefont{Tsumura}},
  \bibinfo{journal}{Phys. Rev. D} \textbf{\bibinfo{volume}{73}},
  \bibinfo{pages}{016006} (\bibinfo{year}{2006}), \eprint{hep-ph/0505191}.

\bibitem[{\citenamefont{Davidson et~al.}(2008)\citenamefont{Davidson, Isidori,
  and Uhlig}}]{Davidson:2007si}
\bibinfo{author}{\bibfnamefont{S.}~\bibnamefont{Davidson}},
  \bibinfo{author}{\bibfnamefont{G.}~\bibnamefont{Isidori}}, \bibnamefont{and}
  \bibinfo{author}{\bibfnamefont{S.}~\bibnamefont{Uhlig}},
  \bibinfo{journal}{Phys. Lett.} \textbf{\bibinfo{volume}{B663}},
  \bibinfo{pages}{73} (\bibinfo{year}{2008}), \eprint{0711.3376}.

\bibitem[{\citenamefont{Moreau and Silva-Marcos}(2006)}]{Moreau:2006np}
\bibinfo{author}{\bibfnamefont{G.}~\bibnamefont{Moreau}} \bibnamefont{and}
  \bibinfo{author}{\bibfnamefont{J.}~\bibnamefont{Silva-Marcos}},
  \bibinfo{journal}{JHEP} \textbf{\bibinfo{volume}{03}}, \bibinfo{pages}{090}
  (\bibinfo{year}{2006}), \eprint{hep-ph/0602155}.

\bibitem[{\citenamefont{Agashe et~al.}(2006)\citenamefont{Agashe, Blechman, and
  Petriello}}]{Agashe:2006iy}
\bibinfo{author}{\bibfnamefont{K.}~\bibnamefont{Agashe}},
  \bibinfo{author}{\bibfnamefont{A.~E.} \bibnamefont{Blechman}},
  \bibnamefont{and}
  \bibinfo{author}{\bibfnamefont{F.}~\bibnamefont{Petriello}},
  \bibinfo{journal}{Phys. Rev. D} \textbf{\bibinfo{volume}{74}},
  \bibinfo{pages}{053011} (\bibinfo{year}{2006}), \eprint{hep-ph/0606021}.

\bibitem[{\citenamefont{Huber}(2003)}]{Huber:2003tu}
\bibinfo{author}{\bibfnamefont{S.~J.} \bibnamefont{Huber}},
  \bibinfo{journal}{Nucl. Phys.} \textbf{\bibinfo{volume}{B666}},
  \bibinfo{pages}{269} (\bibinfo{year}{2003}), \eprint{hep-ph/0303183}.

\bibitem[{\citenamefont{Cirigliano et~al.}(2021)\citenamefont{Cirigliano,
  Fuyuto, Lee, Mereghetti, and Yan}}]{Cirigliano:2021img}
\bibinfo{author}{\bibfnamefont{V.}~\bibnamefont{Cirigliano}},
  \bibinfo{author}{\bibfnamefont{K.}~\bibnamefont{Fuyuto}},
  \bibinfo{author}{\bibfnamefont{C.}~\bibnamefont{Lee}},
  \bibinfo{author}{\bibfnamefont{E.}~\bibnamefont{Mereghetti}},
  \bibnamefont{and} \bibinfo{author}{\bibfnamefont{B.}~\bibnamefont{Yan}},
  \bibinfo{journal}{JHEP} \textbf{\bibinfo{volume}{03}}, \bibinfo{pages}{256}
  (\bibinfo{year}{2021}), \eprint{2102.06176}.

\bibitem[{\citenamefont{De}(2024)}]{De:2024foq}
\bibinfo{author}{\bibfnamefont{B.}~\bibnamefont{De}}, \bibinfo{journal}{Phys.
  Lett. B} \textbf{\bibinfo{volume}{855}}, \bibinfo{pages}{138784}
  (\bibinfo{year}{2024}), \eprint{2405.06970}.

\bibitem[{\citenamefont{Takeuchi et~al.}(2017)\citenamefont{Takeuchi, Uesaka,
  and Yamanaka}}]{Takeuchi:2017btl}
\bibinfo{author}{\bibfnamefont{M.}~\bibnamefont{Takeuchi}},
  \bibinfo{author}{\bibfnamefont{Y.}~\bibnamefont{Uesaka}}, \bibnamefont{and}
  \bibinfo{author}{\bibfnamefont{M.}~\bibnamefont{Yamanaka}},
  \bibinfo{journal}{Phys. Lett.} \textbf{\bibinfo{volume}{B772}},
  \bibinfo{pages}{279} (\bibinfo{year}{2017}), \eprint{1705.01059}.

\bibitem[{\citenamefont{Georgi et~al.}(1978)\citenamefont{Georgi, Glashow,
  Machacek, and Nanopoulos}}]{Georgi:1977gs}
\bibinfo{author}{\bibfnamefont{H.~M.} \bibnamefont{Georgi}},
  \bibinfo{author}{\bibfnamefont{S.~L.} \bibnamefont{Glashow}},
  \bibinfo{author}{\bibfnamefont{M.~E.} \bibnamefont{Machacek}},
  \bibnamefont{and} \bibinfo{author}{\bibfnamefont{D.~V.}
  \bibnamefont{Nanopoulos}}, \bibinfo{journal}{Phys. Rev. Lett.}
  \textbf{\bibinfo{volume}{40}}, \bibinfo{pages}{692} (\bibinfo{year}{1978}).

\bibitem[{\citenamefont{Spira et~al.}(1995)\citenamefont{Spira, Djouadi,
  Graudenz, and Zerwas}}]{Spira:1995rr}
\bibinfo{author}{\bibfnamefont{M.}~\bibnamefont{Spira}},
  \bibinfo{author}{\bibfnamefont{A.}~\bibnamefont{Djouadi}},
  \bibinfo{author}{\bibfnamefont{D.}~\bibnamefont{Graudenz}}, \bibnamefont{and}
  \bibinfo{author}{\bibfnamefont{P.~M.} \bibnamefont{Zerwas}},
  \bibinfo{journal}{Nucl. Phys. B} \textbf{\bibinfo{volume}{453}},
  \bibinfo{pages}{17} (\bibinfo{year}{1995}), \eprint{hep-ph/9504378}.

\bibitem[{\citenamefont{Zyla et~al.}(2020)}]{Zyla:2020zbs}
\bibinfo{author}{\bibfnamefont{P.}~\bibnamefont{Zyla}} \bibnamefont{et~al.}
  (\bibinfo{collaboration}{Particle Data Group}), \bibinfo{journal}{PTEP}
  \textbf{\bibinfo{volume}{2020}}, \bibinfo{pages}{083C01}
  (\bibinfo{year}{2020}).

\bibitem[{\citenamefont{Altmannshofer et~al.}(2019)}]{Belle-II:2018jsg}
\bibinfo{author}{\bibfnamefont{W.}~\bibnamefont{Altmannshofer}}
  \bibnamefont{et~al.} (\bibinfo{collaboration}{Belle-II}),
  \bibinfo{journal}{PTEP} \textbf{\bibinfo{volume}{2019}},
  \bibinfo{pages}{123C01} (\bibinfo{year}{2019}), \bibinfo{note}{[Erratum: PTEP
  2020, 029201 (2020)]}, \eprint{1808.10567}.

\bibitem[{\citenamefont{Celis et~al.}(2014)\citenamefont{Celis, Cirigliano, and
  Passemar}}]{Celis:2014asa}
\bibinfo{author}{\bibfnamefont{A.}~\bibnamefont{Celis}},
  \bibinfo{author}{\bibfnamefont{V.}~\bibnamefont{Cirigliano}},
  \bibnamefont{and} \bibinfo{author}{\bibfnamefont{E.}~\bibnamefont{Passemar}},
  \bibinfo{journal}{Phys. Rev.} \textbf{\bibinfo{volume}{D89}},
  \bibinfo{pages}{095014} (\bibinfo{year}{2014}), \eprint{1403.5781}.

\bibitem[{\citenamefont{Dulat et~al.}(2016)\citenamefont{Dulat, Hou, Gao,
  Guzzi, Huston, Nadolsky, Pumplin, Schmidt, Stump, and Yuan}}]{Dulat:2015mca}
\bibinfo{author}{\bibfnamefont{S.}~\bibnamefont{Dulat}},
  \bibinfo{author}{\bibfnamefont{T.-J.} \bibnamefont{Hou}},
  \bibinfo{author}{\bibfnamefont{J.}~\bibnamefont{Gao}},
  \bibinfo{author}{\bibfnamefont{M.}~\bibnamefont{Guzzi}},
  \bibinfo{author}{\bibfnamefont{J.}~\bibnamefont{Huston}},
  \bibinfo{author}{\bibfnamefont{P.}~\bibnamefont{Nadolsky}},
  \bibinfo{author}{\bibfnamefont{J.}~\bibnamefont{Pumplin}},
  \bibinfo{author}{\bibfnamefont{C.}~\bibnamefont{Schmidt}},
  \bibinfo{author}{\bibfnamefont{D.}~\bibnamefont{Stump}}, \bibnamefont{and}
  \bibinfo{author}{\bibfnamefont{C.~P.} \bibnamefont{Yuan}},
  \bibinfo{journal}{Phys. Rev. D} \textbf{\bibinfo{volume}{93}},
  \bibinfo{pages}{033006} (\bibinfo{year}{2016}), \eprint{1506.07443}.

\end{thebibliography}

\end{document}